\documentclass[reprint,twocolumn,showpacs,preprintnumbers,amsmath,amssymb,pra,aps,superscriptaddress]{revtex4-1}

\usepackage{graphicx}
\usepackage{dcolumn}
\usepackage{bm}
\usepackage{dcolumn}
\usepackage{amsmath}
\usepackage{amssymb}
\usepackage{longtable}
\usepackage{array}
\usepackage[version=3]{mhchem}
\usepackage{epic}
\usepackage{epstopdf}
\usepackage{verbatim}

\usepackage[usenames,dvipsnames,svgnames,table]{xcolor}

\newcolumntype{C}[1]{>{\centering}m{#1}}
\newcommand{\etal}{\textit{et~al.}}

\newcommand{\muB}{~\mu _\text{B}}
\newcommand{\TmuB}{\text{T} / \mu _\text{B}}
\begin{document}

\title{Study of the hyperfine fields in the \boldmath BaFe$_2$As$_2$\unboldmath ~family \\and its relation to the magnetic moment}  
\author{Gerald Derondeau}    \email{gerald.derondeau@cup.uni-muenchen.de}
\affiliation{%
  Department  Chemie,  Physikalische  Chemie,  Universit\"at  M\"unchen,
  Butenandtstr.  5-13, 81377 M\"unchen, Germany\\}

\author{J\'an Min\'ar}
\affiliation{%
  Department  Chemie,  Physikalische  Chemie,  Universit\"at  M\"unchen,
  Butenandtstr. 5-13, 81377 M\"unchen, Germany\\}
\affiliation{%
  NewTechnologies-Research Center, University of West Bohemia, Pilsen, Czech Republic\\}

\author{Hubert Ebert}
\affiliation{%
  Department  Chemie,  Physikalische  Chemie,  Universit\"at  M\"unchen,
  Butenandtstr. 5-13, 81377 M\"unchen, Germany\\}

\date{\today}


\begin{abstract}
The hyperfine field $B_\text{hf}$ and the magnetic properties of the BaFe$_2$As$_2$ family are studied using the fully relativistic Dirac formalism for different types of substitution. The study covers electron doped Ba(Fe$_{1-x}$Co$_x$)$_2$As$_2$ and Ba(Fe$_{1-x}$Ni$_x$)$_2$As$_2$, hole doped (Ba$_{1-x}$K$_x$)Fe$_2$As$_2$ and also isovalently doped Ba(Fe$_{1-x}$Ru$_x$)$_2$As$_2$ and BaFe$_2$(As$_{1-x}$P$_x$)$_2$ for a wide range of the concentration $x$. 
For the substituted compounds the hyperfine fields show a very strong dependence on the dopant type and its concentration $x$.
Relativistic contributions were found to have a significantly stronger impact for the iron pnictides when compared to bulk Fe. As an important finding, we demonstrate that it is not sensible to relate the hyperfine field $B_\text{hf}$ to the average magnetic moment $\mu$ of the compound, as it was done in earlier literature.

\end{abstract}
                
\maketitle


\section{Introduction}
Since the discovery of high-temperature superconductivity in La(O$_{1-x}$F$_x$)FeAs \cite{KWHH08,TIA+08} the iron pnictides are currently one of the most important prototype systems for unconventional superconductivity. The mechanism of superconductivity is more than likely connected to magnetic fluctuations \cite{Sin08a,MSJD08,FPT+10}, which makes the magnetic behavior of the iron pnictides a crucial property \cite{MJ09, MJB+08}. Despite tremendous research over the last years the complex magnetism of these compounds is still non-trivial to explain and some problems remain unsolved.

For example, a discrepancy is observed concerning the magnitude of the magnetic moment, depending on the chosen experimental method. Neutron diffraction data predicts for the low-temperature phase of BaFe$_2$As$_2$ a total magnetic moment of $0.87\muB$ per Fe from powder samples \cite{HQB+08}, while from \ce{^{57}Fe}~M\"ossbauer spectroscopy \cite{RTJ+08,RTS+09} a value between $0.4$ and $0.5\muB$ was estimated. One should note that the magnetic moments in the iron pnictides are generally considered to behave nearly itinerant \cite{MJ09,MSJD08,YLAA09,OKZ+09,FTO+09}, although sometimes a localized picture might be more appropriate \cite{HYPS09,YZO+09,CLLR08}. Furthermore, density functional theory (DFT) calculations often  overestimate the magnitude of the magnetic moments, ranging from approximately $1.2\muB$ up to $2.6\muB$ \cite{KOK+09,YLAA09,SBP+11,MJB+08,AC09}. Thus, the magnetic moments are known to be highly sensitive to the system and computational parameters, which makes estimations difficult and leads sometimes to seemingly contradicting reports \cite{MJB+08,SLS+09,RTJ+08,GLK+11,VFB+12}. Furthermore, the importance of spin-orbit coupling for the iron pnictides was only recently stressed \cite{BEL+16}.

Nowadays, a lot of \ce{^{57}Fe}~M\"ossbauer spectroscopy data are available for the BaFe$_2$As$_2$ family with different types of substitution and doping \cite{RTS+09,BRCF10,NFN+10,RBG+14}. The previously mentioned discrepancy between neutron diffraction and \ce{^{57}Fe}~M\"ossbauer spectroscopy is often ascribed to possible non-zero contributions of $d$-orbitals to the hyperfine field with an opposite sign to that of the Fermi contact field \cite{BRCF10}. This would explain why the suggested hyperfine proportionality constant $A$ between the experimentally measured hyperfine field $B_\text{exp}$ and the underlying magnetic moment $\mu$(Fe) has a non-linear behavior and is in particular not comparable to the corresponding value for bulk Fe. This would imply that a more reliable estimation of magnetic moments based on \ce{^{57}Fe}~M\"ossbauer spectroscopy lies not between $0.4$ and $0.5\muB$ but has a higher value. Although such aspects were already suggested as a most likely explanation for this discrepancy \cite{BRCF10}, a quantitative study of the theoretical hyperfine fields including relativistic contributions is still missing \cite{PFL11}.

To clarify this situation, we address in this paper the antiferromagnetic state of the undoped mother compound BaFe$_2$As$_2$ together with a large variety of different types of substitution. These include electron doping in the case of Ba(Fe$_{1-x}$Co$_x$)$_2$As$_2$ and Ba(Fe$_{1-x}$Ni$_x$)$_2$As$_2$, hole doping as in (Ba$_{1-x}$K$_x$)Fe$_2$As$_2$ and also isovalently doped compounds like Ba(Fe$_{1-x}$Ru$_x$)$_2$As$_2$ and BaFe$_2$(As$_{1-x}$P$_x$)$_2$. To deal adequately with substitutional systems the fully relativistic Korringa-Kohn-Rostoker-Green function (KKR-GF) approach is used, which was already shown to be an appropriate tool to investigate various properties of the iron pnictides \cite{DPM+14,DBEM16,DBB+16}. Chemical disorder due to substitution is dealt by means of the coherent potential approximation (CPA), which effectively gives results comparable to the tedious average over many supercell configurations and is much more reliable than the virtual crystal approximation (VCA) \cite{DPM+14,BLGK12}. Application of the CPA to the iron pnictides was already shown to be quite successful \cite{KJ14,KAJ14,DPM+14,HHS15,DBEM16}. Using this approach, one cannot only investigate the type-resolved evolution of magnetic moments with composition, but also the doping dependence of the hyperfine fields. Furthermore, all contributions to the total hyperfine field $B_\text{hf}$ can be separated, revealing the direct impact of orbital non-$s$-electron parts within the fully relativistic approach.
  
%

\section{Computational details}
All calculations have been performed self-consistently and fully relativistically within the four component Dirac formalism, using the Munich SPR-KKR program package \cite{EKM11,SPR-KKR6.3_2}.
The crystal structure is based on the orthorhombic, antiferromagnetic phase of BaFe$_2$As$_2$ in its experimentally observed stripe spin state using a 4-Fe unit cell. This implies antiferromagnetically ordered chains along the $a$ and $c$ axes and ferromagnetically ordered chains along the $b$ axis. The lattice parameters and As position $z$ where chosen according to experimental X-ray data \cite{RTJ+08}. To account for the influence of different substitutions, a linear interpolation of the lattice parameters with respect to the concentration $x$ was performed based on Vegard's law \cite{Veg21}. This interpolation was individually done for each type of substitution, based on available experimental data \cite{RTJ+08,SJM+08,RPTJ08,TNK+10,Rot10,MSN+16}. More details on this procedure can be found in previous publications \cite{DPM+14,DBEM16}. The treatment of disorder introduced by substitution is dealt with by means of the CPA. For the angular momentum expansion of the KKR Green function an upper limit $\ell_\text{max} = 4$ was used, i.e. $s$, $p$, $d$, $f$ and $g$ orbitals were included in the basis set, although contributions to the hyperfine field of Fe from $f$ and $g$ orbitals are zero as one would expect. All DFT calculations used the local spin density approximation (LSDA) exchange-correlation potential with the parameterization as given by Vosko, Wilk and Nusair \cite{VWN80}. 
The calculation and decomposition of the hyperfine field $B_\text{hf}$ is done in its fully relativistic form as discussed in detail in Ref. \cite{BE01}. 


%
\section{Results and Discussion}

\subsection{Undoped mother compound}
The calculated total magnetic moment of Fe in the undoped mother compound BaFe$_2$As$_2$ is $\mu$(Fe) = 0.73$\muB$, as was already published in earlier work \cite{DBEM16}. This moment splits into a spin magnetic moment of $\mu_\text{spin}$(Fe) = 0.70$\muB$ and an orbital magnetic moment of $\mu_\text{orb}$(Fe) = 0.03$\muB$. Obviously, this is in good agreement with experimental neutron diffraction data of pure BaFe$_2$As$_2$ being 0.87$\muB$ \cite{HQB+08}.

If the finite size of the atomic core is ignored, as usually done, the fully relativistic approach described in Ref. \cite{BE01} splits $B_\text{hf}$ into five contributions. 
There are two contributions due to the $s$-electrons that are conventionally ascribed to the Fermi contact interaction. The larger part is the core polarization contribution $B_s^c$ that was demonstrated in numerous studies to be proportional to the local spin magnetic moment $\mu_\text{spin}$ \cite{AAB+84,EEG86,LS88}. In addition, there is a $s$-electron contribution from the valence band $B_s^v$ that is due to the polarization and also dominantly due to the population mechanism \cite{EWJP89}. For systems with low symmetry there may be a spin dipolar contribution to $B_\text{hf}$ for the non-$s$ electrons \cite{GE96,BE01}. Apart from $p_{1/2}$-states, states with higher angular momentum like $p$- and $d$-states have zero probability density at the core and for that reason do not contribute to $B_\text{hf}$ via the Fermi contact term. If spin-orbit coupling is accounted for, as done here, there is an additional contribution due to the spin-orbit induced orbital magnetization \cite{GE96,BE01}. As the orbital contribution is in general dominating compared to the spin-dipolar one \cite{BE01} we use in the following the term orbital for the total field connected with non-$s$ electrons. Thus, for a transition metal the remaining three contributions are the orbital field $B_{ns}^c$ of the non-$s$ core states and the orbital fields  $B_p^v$ and $B_d^v$ of the valence electrons with $p$- and $d$-character, respectively. With this one arrives for the hyperfine field $B_\text{hf}$ at the following decomposition \cite{BE01}:

\begin{equation}
 B_\text{hf} = B_s^c + B_s^v + B_{ns}^c + B_p^v + B_d^v\;,\label{eq_Bhf}
\end{equation}

\begin{figure}[tb]
{\includegraphics[clip]{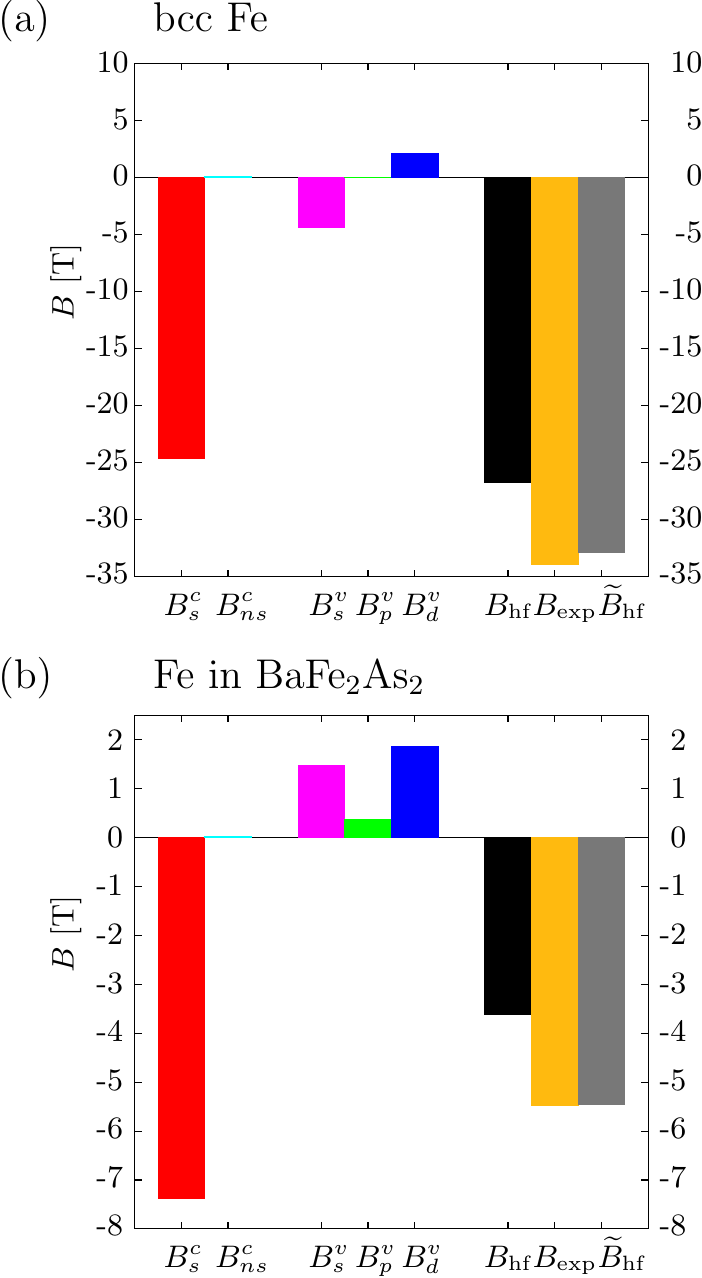}} 
\caption{Contributions to the hyperfine field ${B}_\text{hf}$ for (a) bcc Fe and for (b) Fe in antiferromagnetic BaFe$_2$As$_2$. For comparison experimental values are shown as $B_\text{exp}$ \cite{Dub09,RTJ+08,RTS+09}. $\widetilde{B}_\text{hf}$  is based on Eq. \eqref{eq_Bhftilde} and includes an enhancement of the core polarization $B_s^c$ of 25\%.\vspace*{-0.2cm}}\label{Fig_1}
\end{figure}

Fig.~\ref{Fig_1} (a) shows for bcc Fe numerical results for the various contributions to the hyperfine field. As it is well known, $B_\text{hf}$ of bcc Fe is dominated by its large core polarization contribution $B^c_s$. This is enhanced by the field $B^v_s$ which is also negative. All other contributions are much smaller and positive. Comparing the total calculated hyperfine field $B_\text{hf} = -26.7$\;T with the corresponding experimental value $B_\text{exp} = -33.9$\;T  one finds the theoretical values too small by about 25\%. This well known problem is primarily to be ascribed to shortcomings of LSDA when dealing with the core polarization cased by the spin polarization of the valence electrons \cite{BAZD87,ESG88,BIM+11}. To cure this problem it is common to enhance $B^c_s$ by about 25\% \cite{BAZD87,ESG88,EA93,BIM+11}. Using this empirical approach one has for the enhanced hyperfine field $\widetilde{B}_\text{hf}$ the relation \eqref{eq_Bhftilde}:

\begin{equation}
 \widetilde{B}_\text{hf} = 1.25 \cdot B_s^c + B_s^v + B_{ns}^c + B_p^v + B_d^v\;.\label{eq_Bhftilde}
\end{equation}

As can be seen in Fig.~\ref{Fig_1} (a) this leads to $\widetilde{B}_\text{hf} = -32.9$\;T for bcc Fe, in good agreement with experiment.  
Next, consider Fe in the undoped mother compound BaFe$_2$As$_2$ as presented in Fig.~\ref{Fig_1} (b). Comparing the calculated $B_\text{hf} = -3.62$\;T with the experimental one $B_\text{exp} = -5.47$\;T \cite{RTJ+08,RTS+09} the shortcomings of LSDA are obviously the same as for bcc Fe as one would expect. However, the enhanced field $\widetilde{B}_\text{hf} = -5.46$\;T is in perfect agreement with experiment, confirming the transferability of the enhancement factor in Eq.~\eqref{eq_Bhftilde}.
Compared with bcc Fe the various contributions to $\widetilde{B}_\text{hf}$ of Fe in BaFe$_2$As$_2$ show two major differences.
First, the sign of the valence band $s$-electrons contribution $B_s^v$ is different and second, the spin-orbit induced contribution of $d$-electrons $B_d^v$ is considerably higher in the later case. 
Both features lead to a very different relation between the enhanced hyperfine field $\widetilde{B}_\text{hf}$ and the local spin magnetic moment $\mu_\text{spin}$ for the two systems. As $\widetilde{B}_\text{hf}$ of bcc Fe is dominated by its enhanced core polarization contribution $\widetilde{B}^c_s$ ($\widetilde{B}_\text{hf} / \widetilde{B}^c_s \approx 1.07$), which is proportional to $\mu_\text{spin}$, it seems justified to assume that the experimental field $B_\text{exp}$ reflects in a one-to-one manner the local spin moment. For Fe in BaFe$_2$As$_2$, on the other hand, we find $\widetilde{B}_\text{hf} / \widetilde{B}^c_s \approx 0.59$, i.e. the total field $\widetilde{B}_\text{hf}$ can by no means be used to monitor the local spin magnetic moment of Fe.

\subsection{Electron and hole doping}
Having investigated the hyperfine field contributions of the undoped BaFe$_2$As$_2$ including relativistic effects, an interesting issue is their variation under different types of substitution in the BaFe$_2$As$_2$ family. 

\begin{figure}[tbh]
{\includegraphics[clip]{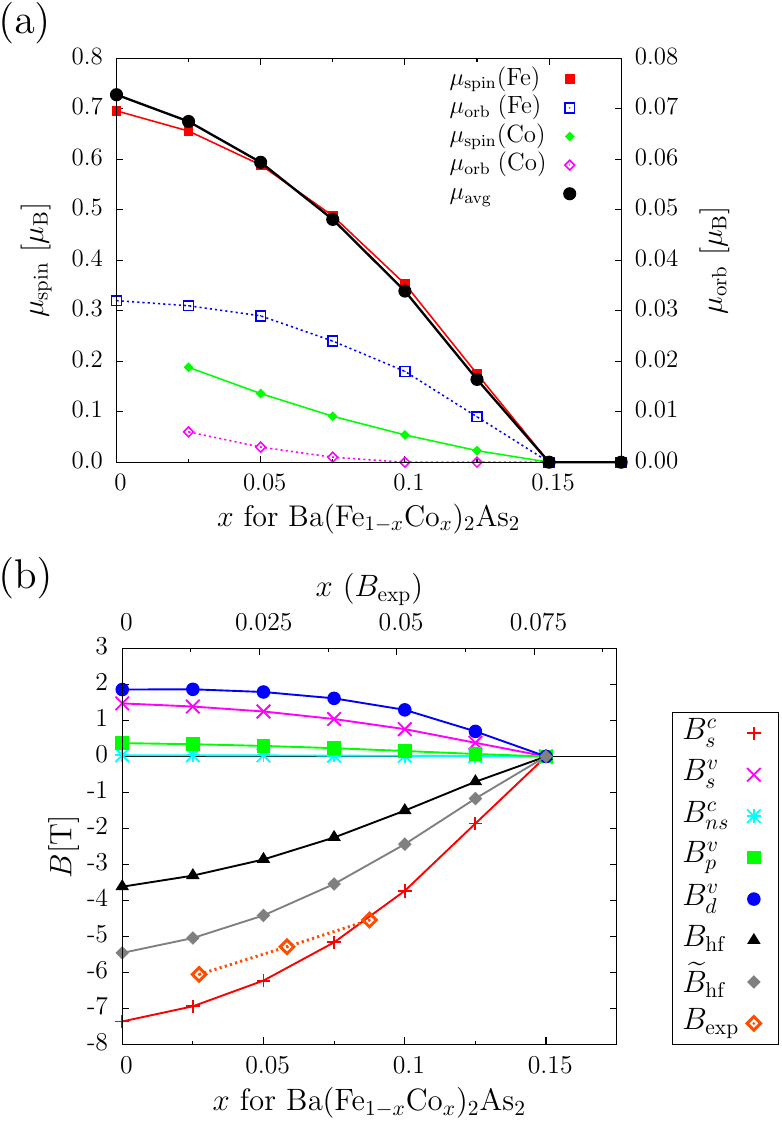}} 
\caption{(a) Component-resolved magnetic moments for Co-122 depending on the concentration $x$. The left (right) scale refers to the spin (orbital) magnetic moment. (b) Corresponding hyperfine field contributions for Fe in Co-122. The experimental data $B_\text{exp}$ (dashed orange lines) \cite{BRCF10} refer to the upper axis, with the upper and lower axes for the concentration $x$ chosen such that $x_\text{crit} = x_\text{crit,exp}$. \vspace*{-0.2cm}}\label{Fig_Co}
\end{figure}

\begin{figure}[tbh]
{\includegraphics[clip]{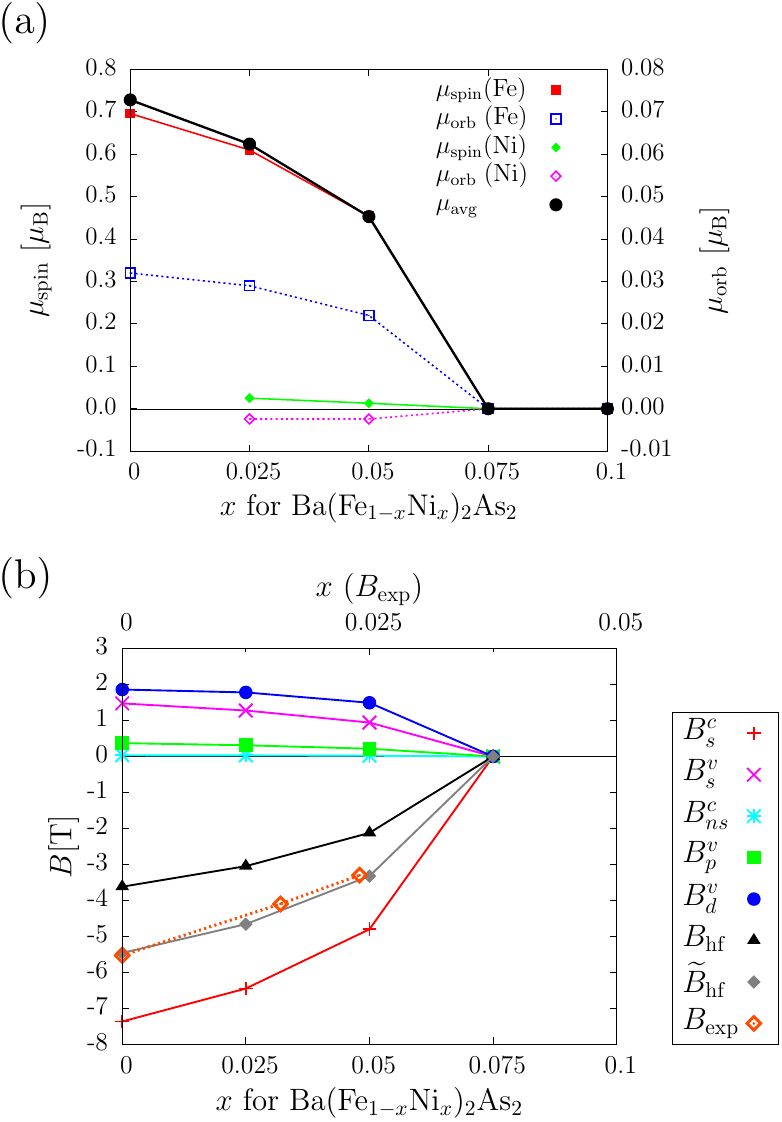}} 
\caption{Same as for Fig.~\ref{Fig_Co}, but for Ni-122 with experimental data from Ref. \cite{NFN+10}. \vspace*{-0.2cm}}\label{Fig_Ni}
\end{figure}

\begin{figure}[tbh]
{\includegraphics[clip]{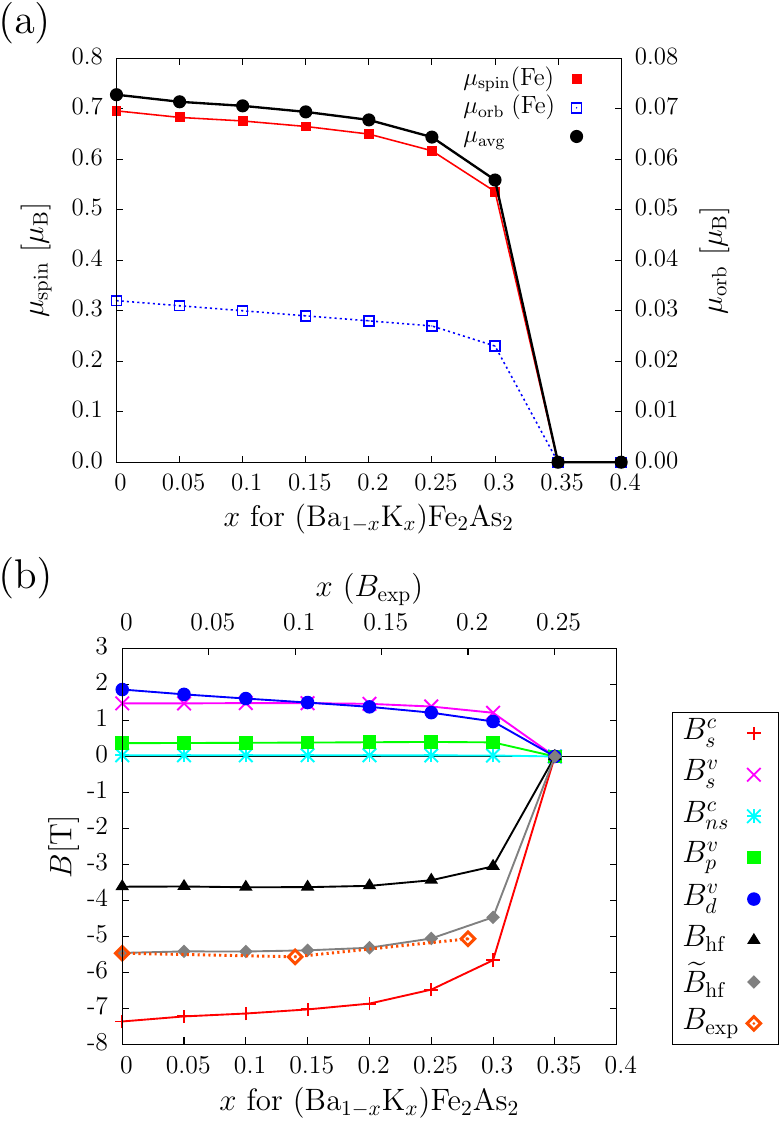}} 
\caption{Same as for Fig.~\ref{Fig_Co}, but for K-122 with experimental data from Ref. \cite{RTS+09}. \vspace*{-0.2cm}}\label{Fig_K}
\end{figure}

Two examples of electron doping were investigated, namely Ba(Fe$_{1-x}$Co$_x$)$_2$As$_2$ (Co-122) and Ba(Fe$_{1-x}$Ni$_x$)$_2$As$_2$ (Ni-122), with the corresponding data shown in Fig.~\ref{Fig_Co} and \ref{Fig_Ni}, respectively. Furthermore, one case of hole doping, (Ba$_{1-x}$K$_x$)Fe$_2$As$_2$ (K-122) has been considered (see Fig.~\ref{Fig_K}). In all cases, the magnetic moments of the components are presented in panel (a) as a function of the concentration. The magnetic moments for Co-122 in Fig.~\ref{Fig_Co} (a) were published before \cite{DBEM16}, and are reproduced here to supply a reference for the hyperfine field and to allow for direct comparison with other systems. The various figures give in a component-resolved manner the spin magnetic moments $\mu_\text{spin}$ (left axis) and the orbital magnetic moments $\mu_\text{orb}$ (right axis). The concentration dependent average of the system with composition Ba(Fe$_{1-x}$\textit{TM}$_x$)$_2$As$_2$ is shown as $\mu_\text{avg} = (1 -x) \cdot [\mu_\text{spin}\text{(Fe)} + \mu_\text{orb}\text{(Fe)}] + x \cdot [\mu_\text{spin}\text{(\textit{TM})} + \mu_\text{orb}\text{(\textit{TM})}]$. 

First consider the electron doped compounds Co-122 and Ni-122. Both systems show a similar decrease in $\mu_\text{avg}$ until the breakdown of long range antiferromagnetic (AFM) order at $x_\text{crit}$ is reached, with $x_\text{crit}\text{(Co-122)} = 0.125$ and $x_\text{crit}\text{(Ni-122)} = 0.075$, respectively. 
This is in reasonable agreement with experiment, with the experimental $x_\text{crit,exp}$ being lower ($x_\text{crit,exp}\text{(Co-122)} \approx 0.075$ \cite{LCA+09}, $x_\text{crit,exp}\text{(Ni-122)} \approx 0.0375$ \cite{NTY+10}). Concerning the instability of the antiferromagnetic order, the electronic structure calculations account for a change in the nesting condition due to a shift of the Fermi level due to doping but they do not explicitly account for fluctuating magnetic moments or incommensurate spin-density waves \cite{PKK+11}. This might explain the observed discrepancies between $x_\text{crit}$ and $x_\text{crit,exp}$, implying that these aspects should be accounted for in order to get better agreement.

In line with experiment, $x_\text{crit}$ for Ni-122 is found to be only half of Co-122. This had to be expected because of the formal doubling of electron doping by Ni compared to Co substitution of Fe. Another difference between these two compounds is the lower Ni moment in Ni-122 compared to that of Co in Co-122. In this context one should also note that the rather small orbital moment of Ni has a different sign compared to its spin moment. 
The various hyperfine field contributions for Fe in Co-122 and Ni-122 are shown in Fig.~\ref{Fig_Co} (b) and Fig.~\ref{Fig_Ni} (b), respectively. The trends of the Fe magnetic moments and in the hyperfine field contributions behave in a similar way. 
The figures show also experimental data for the hyperfine field $B_\text{exp}$ \cite{BRCF10,NFN+10}. These has been plotted using a different scale for the concentration $x$ at the top of the figure that was chosen such that theoretical and experimental critical concentrations agree ($x_\text{crit} = x_\text{crit,exp}$). With the afore mentioned enhancement of the core polarization field by 25\% and the rescaling of the $x$-axis one finds a very satisfying agreement for $\widetilde{B}_\text{hf}$ and $B_\text{exp}$ for Co-122 (Fig.~\ref{Fig_Co} (b)) as well as Ni-122 (Fig.~\ref{Fig_Ni} (b)).

Next, the K-122 compound is discussed with its magnetic moments shown in Fig.~\ref{Fig_K} (a) (see also Ref. \cite{DMWE16}). A breakdown of the AFM order is found from the calculations at $x_\text{crit}\text{(K-122)} = 0.35$, while a lower $x_\text{crit,exp}\text{(K-122)} \approx 0.25$ \cite{RPTJ08} is observed in experiment. It should be noted that the substituted K does not have a noteworthy magnetic moment. As the Fe concentration does not change with substitution on the Ba position, the average moment is therefore equal to the Fe moment, leading in this case to $\mu_\text{avg} = \mu\text{(Fe)} = \mu_\text{spin}\text{(Fe)} + \mu_\text{orb}\text{(Fe)}$. One can see that for K-122 the magnetic moments change only marginally over a wide concentration range $x$ and undergo a sharp drop for $x > 0.25$. The same behavior can be seen in the hyperfine field contributions of K-122 as shown in Fig.~\ref{Fig_K} (b). Experimental data for $B_\text{exp}$ \cite{RTS+09}, referring again to the upper axis, are in good agreement with the enhanced theoretical field $\widetilde{B}_\text{hf}$. In particular, the experimental $B_\text{exp}$ is also nearly constant over a large range of concentration; in variance to the electron doped systems considered above.

\subsection{Isovalent doping}
The subsequently discussed Ba(Fe$_{1-x}$Ru$_x$)$_2$As$_2$ (Ru-122) and BaFe$_2$(As$_{1-x}$P$_x$)$_2$ (P-122) compounds are fundamentally different from the systems considered above because of the isovalent doping. This means in particular, that the VCA is inappropriate to deal with these systems in a meaningful way. Still, a supercell approach could be applied to deal with the substitution \cite{WBW+13}. However, the large computational effort makes theoretical work on these compounds rare and difficult. On the other hand, CPA based approaches provide an efficient and powerful framework for this task. 

\begin{figure}[tbh]
{\includegraphics[clip]{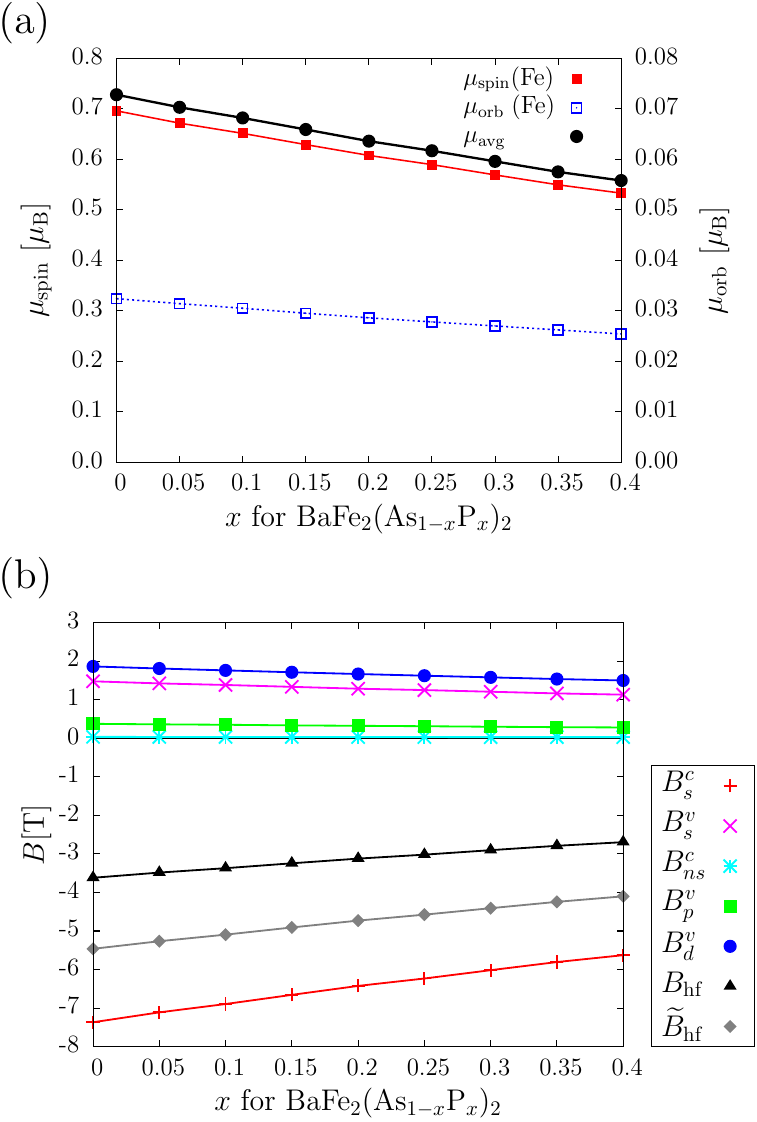}} 
\caption{Same as for Fig.~\ref{Fig_Co}, but for P-122. \vspace*{-0.2cm}}\label{Fig_P}
\end{figure}

\begin{figure}[tbh]
{\includegraphics[clip]{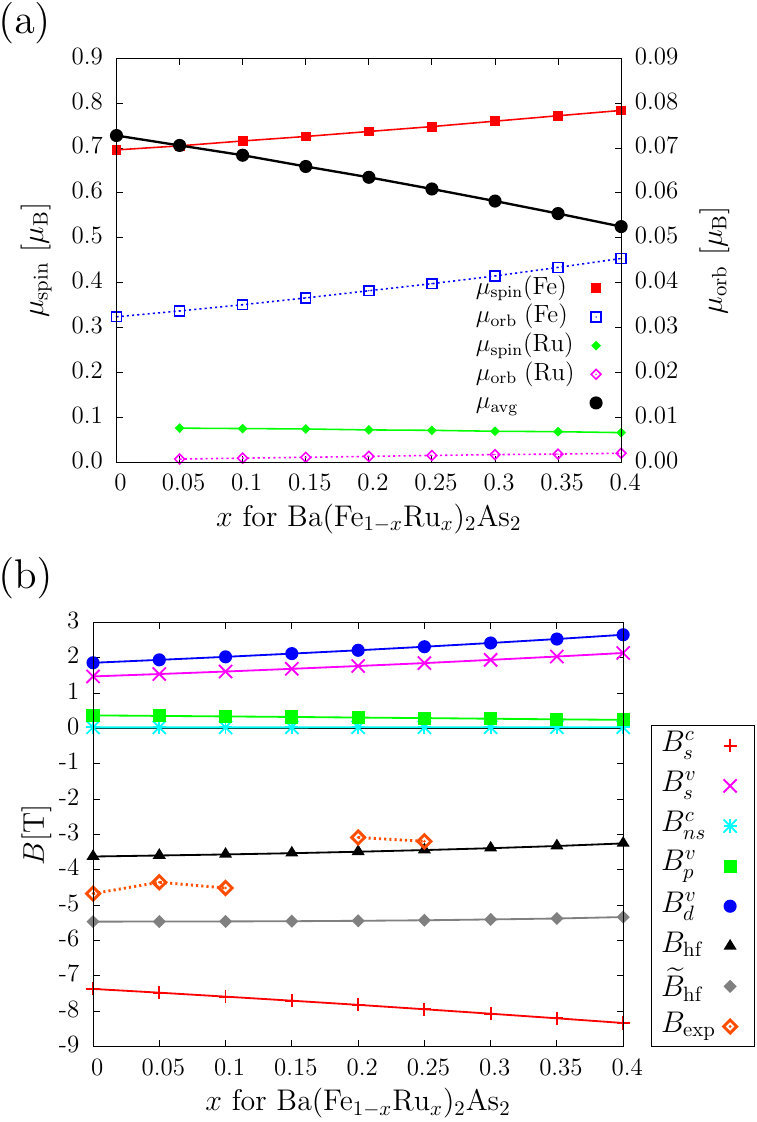}} 
\caption{Same as for Fig.~\ref{Fig_Co}, but for Ru-122 with experimental data from Ref. \cite{RBG+14}. \vspace*{-0.2cm}}\label{Fig_Ru}
\end{figure}

We show the component-resolved magnetic moments of P-122 and Ru-122 in Fig.~\ref{Fig_P} (a) and Fig.~\ref{Fig_Ru} (a), respectively. The first point to note is, that the calculations do not lead to a critical concentration $x_\text{crit}$ within the investigated regime of substitution, while on the experimental side one has $x_\text{crit,exp}\text{(Ru-122)} \approx x_\text{crit,exp}\text{(P-122)} \approx 0.3$ \cite{KPR+11,HLZ+15}. Isovalent doping should in general shift the Fermi level $E_F$ only marginally, leading to an unchanged nesting behavior. Thus, magnetic ordering may be preserved as long as the substitutional limit $x \rightarrow 1$ has a finite magnetic moment. In the case of electron or hole doping  of BaFe$_2$As$_2$ the breakdown of magnetic order at a critical concentration $x_\text{crit}$ can be understood solely by the nesting condition when the Fermi energy $E_F$ changes due to substitution. Note that also K-122 shows a finite $x_\text{crit}$ with good agreement to experiment, although the substitution happens not on the Fe position but within the Ba layer. On the other hand, isovalent substitution either within (Ru-122) or outside (P-122) the Fe layer cannot explain the magnetic breakdown by the substitution alone. This indicates that other phenomena not accounted for within the CPA mean field approach influence the stability of the magnetic structure. In the literature e.g. magnetic dilution was discussed as the main driving force for the magnetic breakdown in Ru-122 \cite{DLF+11,DHR+13}. Although we find a decrease in the magnetic moments due to the decrease in the Fe content, it seems not sufficient to cause a breakdown of the magnetic order without further reasons. Spin fluctuations and incommensurate spin-density waves can have an impact on the stability of the antiferromagnetic order, but also the emergence of a competing superconducting state might play a role. In any case, it becomes obvious that the isovalently doped compounds of the BaFe$_2$As$_2$ family are even more difficult to understand than the electron and hole doped variants. Nevertheless, LSDA-based calculations can reproduce the decrease of the average magnetic moment $\mu_\text{avg}$ for Ru-122 as well as for P-122, although the details of this reduction in the magnetic moments are fundamentally different.

The magnetic moments and the hyperfine field contributions of Fe in P-122 shown in Fig.~\ref{Fig_P} behave in a similar way as those of K-122 (Fig.~\ref{Fig_K}). In both cases the substitution takes place outside the Fe layer; i.e. although the Fe concentration does not change the total Fe moment $\mu\text{(Fe)}$ does. The hyperfine field contributions of Fe in P-122 vary again similar with composition as the magnetic moments do. Of course, this has to be expected as the hyperfine field reflects the magnetization of the Fe atoms, which are the only magnetic components of these systems.

For Ru-122 the average moment $\mu_\text{avg}$ shown in Fig.~\ref{Fig_Ru} (a) decreases due to the increasing concentration of Ru which has a small induced magnetic moment of around $\mu\text{(Ru)} \approx 0.07\muB$, independent on the concentration $x$. However, the local Fe spin magnetic moment $\mu_\text{spin}\text{(Fe)}$ and orbital $\mu_\text{orb}\text{(Fe)}$ magnetic moments surprisingly increase. This is a rather unexpected finding as it was not observed so far within theoretical investigations on the iron pnictides. Accordingly, the corresponding relation to the directly measurable hyperfine field $B_\text{hf}$ of Fe is of interest as it provides an element specific probe of the magnetic properties. As can be seen in Fig.~\ref{Fig_Ru} (b), $B_\text{hf}$ stays more or less constant over the whole investigated regime of substitution, although $\mu\text{(Fe)}$ increases. This is due to the fact that $\mu_\text{spin}\text{(Fe)}$ and $\mu_\text{orb}\text{(Fe)}$ simultaneously increase leading to a subsequent increase of the absolute values of $B_s^c$ and $B_d^v$. Because the sign of both contributions is different, their changes essentially compensate each other. This does not contradict with experimental findings of Reddy \etal~\cite{RBG+14} depicted in Fig.~\ref{Fig_Ru} (b) which show a more or less constant $B_\text{hf}$ for Ru concentrations $x \le 0.1$. The rapid drop to lower $B_\text{hf}$ values for Ru-122 for $x \ge 0.2$ is most likely connected to the proximity to the critical concentration $x_\text{crit}$, which could not be reproduced by our LSDA-based calculations. 

In conclusion, a quite unexpected and interesting variation of the magnetic moments and the hyperfine field with the concentration $x$ of the Ru-122 compound was found which is consistent with experimental findings. This shows in particular, that Ru-122 and P-122 differ more from each other with respect to their magnetic properties as one might expect for two isovalently doped pnictides.

\subsection{Relation to the magnetic moment}
Finally, the results can be used to clarify the relation between $B_\text{hf}$ and the average magnetic moment $\mu_\text{avg}$. It is quite common to assume that the ratio $A_\text{hf}^\text{avg} = -B_\text{hf} / \mu_\text{avg}$ or $A_\text{hf} = -B_\text{hf} / \mu_\text{spin}\text{(Fe)}$ is constant and use this value in order to obtain the magnetic moments in related compounds from the Fe hyperfine fields. For example $A_\text{hf}^\text{avg}\text{(Fe)} = 15$\;$\TmuB$ was given for bulk Fe and $A_\text{hf}^\text{avg}\text{(Fe$^{3+}$)} = 11$\;$\TmuB$ for Fe$^{3+}$ ions in Fe$_2$O$_3$ \cite{NFN+10}. These values give for BaFe$_2$As$_2$ with an experimental hyperfine field $B_\text{exp} = -5.47$\;T a magnetic moment $\mu_\text{avg} \sim 0.4$ -- $0.5 \muB$ \cite{RTJ+08,RTS+09}. Later on it was questioned whether these ratios $A_\text{hf}^\text{avg}$ are applicable to the iron pnictides \cite{NFN+10,BRCF10}. In addition, there is general work showing that a scaling of $B_\text{hf}$ with the corresponding magnetic moment $\mu_\text{avg}$ cannot be assumed \emph{a priori} because $A_\text{hf}^\text{avg}$ varies strongly for different materials \cite{Dub09}.
This is in line with our results that can be used to quantify $A_\text{hf}^\text{avg}$. Additionally, the assumption of a constant ratio $A_\text{hf}^\text{avg}$ for doped systems can be disproved, supporting other work \cite{BRCF10} which concludes that $B_\text{hf}$ is indeed not proportional to $\mu_\text{avg}$ for BaFe$_2$As$_2$ based substitutional systems.

\begin{figure}[t!hb]
{\includegraphics[clip]{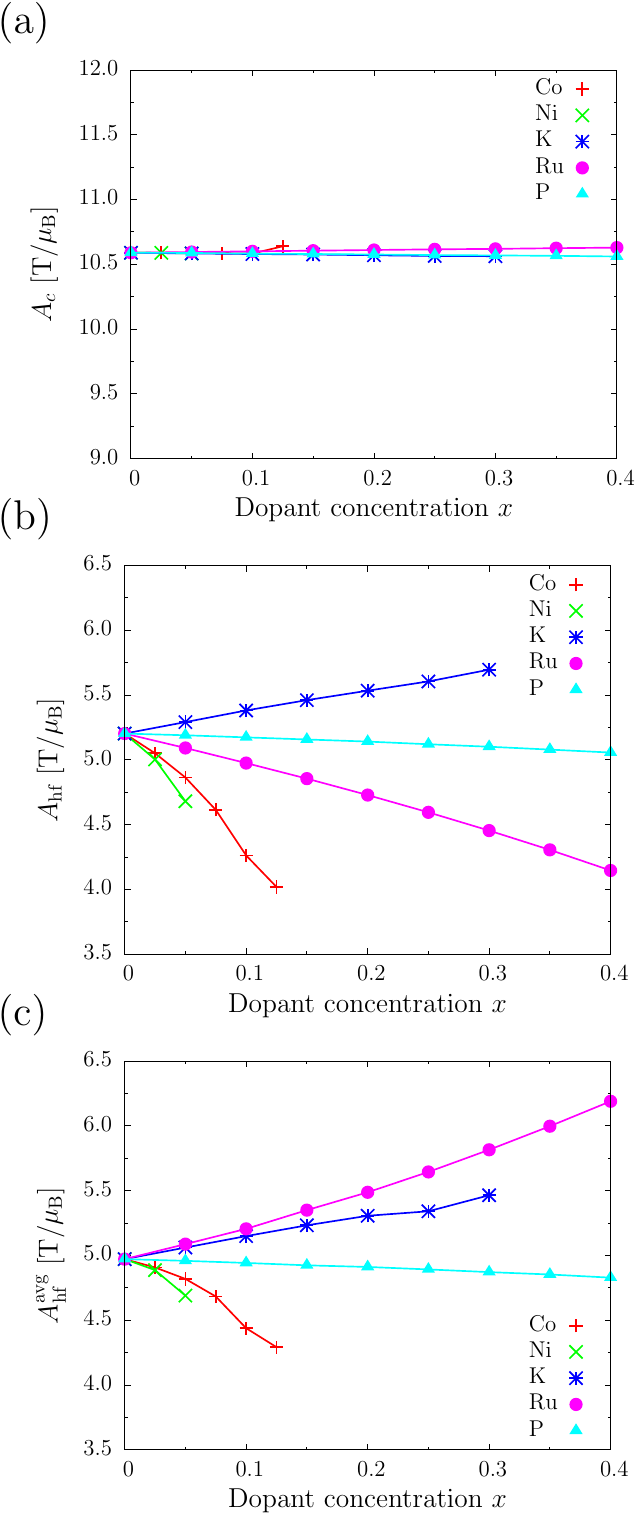}} 
\caption{(a) The ratio $A_c = -B_s^c / \mu_\text{spin}\text{(Fe)}$ is shown for all investigated compounds, depending on the respective dopant and its concentration $x$. The constant behavior shows a reasonable relation to the magnetic moment $\mu$. However, the same ratios are shown for (b) $A_\text{hf} = -B_\text{hf} / \mu_\text{spin}\text{(Fe)}$ and for (c) ${A}_\text{hf}^\text{avg} = -B_\text{hf} / \mu_\text{avg}$, having huge deviations for an constant $A_\text{hf}$ behavior, depending on $x$ and on the chosen dopant. \vspace*{-0.2cm}}\label{Fig_4}
\end{figure}

As stressed already, the core $s$-electron contribution $B_s^c$ is indeed proportional to $\mu_\text{spin}\text{(Fe)}$, which is quantified for our calculations in Fig.~\ref{Fig_4} (a), where we show the ratio $A_c = -B_s^c / \mu_\text{spin}\text{(Fe)}$ for all investigated compounds depending on the concentration $x$. Independent on $x$, we find the value of $A_{c}$ is nearly constant 10.6\;$\TmuB$. This is in reasonable agreement with earlier work of Lindgren and Sj\o str\o m where a value around 12.6\;$\TmuB$ was calculated \cite{LS88}. However, $B_s^c$ can obviously vary significantly from $B_\text{hf}$ as was extensively shown in the literature. 

At least for the undoped BaFe$_2$As$_2$ the average moment equals the total Fe moment and is close to the spin magnetic moment of iron, $\mu_\text{avg} = \mu_\text{spin}\text{(Fe)} + \mu_\text{orb}\text{(Fe)} \approx \mu_\text{spin}\text{(Fe)}$ .
Based on the calculations one gets for BaFe$_2$As$_2$ a ratio $A_\text{hf} = -B_\text{hf} / \mu_\text{spin}\text{(Fe)} = 5.2~\TmuB$ or based on the enhanced hyperfine field $\widetilde{B}_\text{hf}$ a ratio $\widetilde{A}_\text{hf} = 7.8~\TmuB$. This is by a factor of 2 -- 3 different from the ratio $A_\text{hf}^\text{avg}\text{(Fe)}$ applied in previous publications \cite{RTJ+08,RTS+09}. Consequently, the magnetic moment of BaFe$_2$As$_2$ based on the measured hyperfine field of 5.47\;T should be not between $0.4$ and $0.5 \muB$ but rather in the range between $0.7$ -- $1.0 \muB$, which is in better qualitative agreement with neutron diffraction, reporting $0.87 \muB$ \cite{HQB+08}. Nevertheless, one should keep in mind that this is a qualitative estimation and it is clear from the literature \cite{Dub09,BRCF10} and from our work that an estimation of $\mu_\text{avg}$ based on $B_\text{hf}$ should be avoided as far as possible.

However, for the doped iron pnictides there is a significant difference between $\mu_\text{avg}$ and $\mu_\text{spin}\text{(Fe)}$. Thus, the relation between $B_\text{hf}$ and $\mu_\text{avg}$ leads to an unpredictable, non-linear behavior of the ratio $A_\text{hf}^\text{avg}$. To quantify our claim we plot the obtained values of $A_\text{hf} = -B_\text{hf} / \mu_\text{spin}\text{(Fe)}$ and ${A}_\text{hf}^\text{avg} = -B_\text{hf} / \mu_\text{avg}$ depending on the concentration $x$ for all investigated compounds in Fig.~\ref{Fig_4} (b) and (c), respectively. 
Already the ratio $A_\text{hf}$, which is coupled to the spin magnetic moment of Fe, depends strongly on the respective dopant and on the concentration $x$. It becomes apparent that for such a behavior no reasonable relation between $B_\text{hf}$ and $\mu_\text{spin}\text{(Fe)}$ is possible. This problem becomes even more obvious when considering $A_\text{hf}^\text{avg}$. Here, the Ru-122 compound is interesting to mention because $A_\text{hf}$ decreases with $x$ while $A_\text{hf}^\text{avg}$ increases with the concentration. This is due to the fact that the Fe moment in Ru-122 increases while the average moment decreases (see also Fig.~\ref{Fig_Ru}). Thus, it can be crucially misleading to relate $B_\text{hf}$ to the average magnetic moment $\mu_\text{avg}$ in doped iron pnictides.
Consequently, the presented study clearly shows that the hyperfine fields $B_\text{hf}$ of Fe obtained from \ce{^{57}Fe}~M\"ossbauer spectroscopy are not suitable to make predictions about the respective magnetic moment $\mu_\text{avg}$ in doped iron pnictide superconductors for different substitutions.

\section{Summary}

To summarize, this work presented a comprehensive theoretical study of the hyperfine fields in the iron pnictide superconductor family of BaFe$_2$As$_2$ with good agreement with experiment. The CPA was applied to a variety of compounds, dealing accurately with the substitutional disorder and accounting for all variants of doping. This includes electron doped Ba(Fe$_{1-x}$Co$_x$)$_2$As$_2$ and Ba(Fe$_{1-x}$Ni$_x$)$_2$As$_2$, hole doped (Ba$_{1-x}$K$_x$)Fe$_2$As$_2$ and also isovalently doped Ba(Fe$_{1-x}$Ru$_x$)$_2$As$_2$ and BaFe$_2$(As$_{1-x}$P$_x$)$_2$. All systems were investigated in their antiferromagnetic state which was used to study the magnetic moments depending on the concentration $x$ in detail. 
In order to get meaningful results the fully relativistic Dirac formalism was applied, which ensured that all relativistic contributions to $B_\text{hf}$ were accurately dealt with. Indeed, spin-orbit induced contributions were found to have a significantly higher influence on Fe in BaFe$_2$As$_2$ as found for bulk Fe. Consequently, we have quantified in detail why it is not sensible to apply the bulk Fe ratio $A_\text{hf}^\text{avg}\text{(Fe)} = 15$\;$\TmuB$ to the iron pnictides in order to obtain estimations for the magnetic moment from \ce{^{57}Fe}~M\"ossbauer spectroscopy. As a crude estimate, one might rather expect for undoped BaFe$_2$As$_2$ ratios around $5.0$~--~$7.5\;\TmuB$, leading to a magnetic moment of roughly $0.7$ -- $1.0 \muB$ which is more consistent with neutron diffraction reporting $0.87 \muB$ \cite{HQB+08}. However, it is best to avoid such estimations as was shown for the substituted iron pnictide systems. Here, the behavior of $A_\text{hf}^\text{avg}$ with the concentration $x$ is clearly unpredictable and might lead to wrong conclusions. Thus, relating the hyperfine fields $B_\text{hf}$ of Fe obtained via \ce{^{57}Fe}~M\"ossbauer spectroscopy with the magnetic moments should be avoided for substituted iron pnictides. 

%
%
\section*{Acknowledgments}
%
We acknowledge the financial support from the DFG project FOR 1346 and from CENTEM PLUS (L01402).
%



\end{document}